\documentclass{article}
\usepackage{spconf,amsmath,graphicx}
\usepackage{amssymb}
\usepackage{color}
\usepackage[
backend=biber,
style=ieee,
maxbibnames=5,
maxcitenames=5,
doi=false,isbn=false,url=false,eprint=false
]{biblatex} 
\usepackage{setspace}
\defbibheading{bibliography}[\refname]{}

\DeclareSourcemap{
	\maps[datatype=bibtex, overwrite=true]{
		\map{
			\step[fieldsource=booktitle,
			match=\regexp{.*Interspeech.*},
			replace={Proc. Interspeech}]
			\step[fieldsource=journal,
			match=\regexp{.*INTERSPEECH.*},
			replace={Proc. Interspeech}]
			\step[fieldsource=booktitle,
			match=\regexp{.*ICASSP.*},
			replace={Proc. ICASSP}]
			\step[fieldsource=booktitle,
			match=\regexp{.*International.*Conference.*on.*Acoustics,.*Speech.*and.*Signal.*Processing.*},
			replace={Proc. ICASSP}]
			\step[fieldsource=booktitle,
			match=\regexp{.*Multimedia.*and.*Expo.*},
			replace={Proc. ICME}]
			\step[fieldsource=booktitle,
			match=\regexp{.*International.*Conference.*on.*Learning.*Representations.*},
			replace={Proc. ICLR}]
			\step[fieldsource=booktitle,
			match=\regexp{.*International.*Conference.*on.*Machine.*Learning.*},
			replace={Proc. ICML}]
			\step[fieldsource=booktitle,
			match=\regexp{.*Automatic.*Speech.*Recognition.*and.*Understanding.*},
			replace={Proc. ASRU}]
			\step[fieldsource=booktitle,
			match=\regexp{.*Spoken.*Language.*Technology.*},
			replace={Proc. SLT}]
			\step[fieldsource=booktitle,
			match=\regexp{.*Speech.*Synthesis.*Workshop.*},
			replace={Proc. SSW}]
			\step[fieldsource=booktitle,
			match=\regexp{.*workshop.*on.*speech.*synthesis.*},
			replace={Proc. SSW}]
			\step[fieldsource=booktitle,
			match=\regexp{.*Advances.*in.*neural.*information.*processing.*},
			replace={NIPS}]
			\step[fieldsource=booktitle,
			match=\regexp{.*Workshop.*on.* Applications.* of.* Signal.*Processing.*to.*Audio.*and.*Acoustics.*},
			replace={Proc. WASPAA}]
			\step[fieldsource=booktitle,
			match=\regexp{.*International.*Symposium.*on.*Robot.*and.*Human.*Interactive Communication.*},
			replace={Proc. RO-MAN}]
			\step[fieldsource=publisher,
			match=\regexp{.+},
			replace={{}}]
			\step[fieldsource=month,
			match=\regexp{.+},
			replace={{}}]
			\step[fieldsource=location,
			match=\regexp{.+},
			replace={{}}]
			\step[fieldsource=address,
			match=\regexp{.+},
			replace={{}}]
			\step[fieldsource=organization,
			match=\regexp{.+},
			replace={{}}]
		}
	}
}

\title{self-supervised learning method using multiple sampling strategies for general-purpose audio representation}
\name{
   Ibuki Kuroyanagi$^{1,2}$, 
   Tatsuya Komatsu$^{1}$,
}
\address{
   $^1$ LINE Corporation, Tokyo, Japan, $^2$ Nagoya University, Nagoya, Japan
}
\usepackage{amsmath, amssymb}
\begin{document}
%
\maketitle
\begin{abstract}
We propose a self-supervised learning method using multiple sampling strategies to obtain general-purpose audio representation.
Multiple sampling strategies are used in the proposed method to construct contrastive losses from different perspectives and learn representations based on them.
In this study, in addition to the widely used clip-level sampling strategy, we introduce two new strategies, a frame-level strategy and a task-specific strategy.
The proposed multiple strategies improve the performance of frame-level classification and other tasks like pitch detection, which are not the focus of the conventional single clip-level sampling strategy.
We pre-trained the method on a subset of Audioset and applied it to a downstream task with frozen weights.
The proposed method improved clip classification, sound event detection, and pitch detection performance by 25\,\%, 20\,\%, and 3.6\,\%.
\end{abstract}
\section{Introduction}
\label{sec:intro}
Various sound-based applications have been studied, such as speech recognition, speaker identification, and command recognition as speech-related tasks, sound event detection~(SED), pitch detection~(PD), and instrument estimation as non-speech-related tasks. 
These tasks have achieved high performance with the development of neural networks~\cite{zhang2020pushing,chung18b_interspeech,kim21l_interspeech,Komatsu2016,kim_crepe}. 
However, these methods require much effort because they require collecting training data and annotating teacher labels for each task.
For audio tagging tasks, fine-tuning with pre-trained models like PANNs~\cite{kong2020panns}, which are obtained by supervised training with a large dataset such as Audioset~\cite{audioset}, have shown their effectiveness for other audio tagging datasets with small dataset size. 
However, their applicability is limited to tagging tasks because the models are also trained via supervised training for audio tagging. 
Task-specific large datasets are required for obtaining pre-training models for other tasks.
There is a need to establish training methods for general pre-trained models, which can be trained with unlabeled datasets and applied to various tasks.

Self-supervised learning is gaining attention to acquire useful representations for various tasks using unlabeled data.
There are GPT~\cite{NEURIPS2020_1457c0d6} and BERT~\cite{devlin-etal-2019-bert} in natural language processing, CPC~\cite{oord2019representation},  SIMCLR~\cite{Qian_2021_CVPR}, MOCO~\cite{He_2020_CVPR} in the field of computer vision, and CURL~\cite{laskin_srinivas2020curl} in the field of reinforcement learning.
These methods have achieved high performance in their respective fields.
In the field of audio, CPC-based methods design unsupervised loss functions based on regression tasks of representations~\cite{Kharitonov_data,Rivire2020UnsupervisedPT}, and wav2vec2.0 solves a contrast task defined based on the quantization of the learned representation~\cite{Baevski2020vqwav2vec}. 
However, many methods have achieved high performance in speech-related tasks, and few studies have focused on non-speech-related tasks. 
TRILL~\cite{shor20_interspeech} and~\cite{Jansen_2018} are examples of research to obtain general-purpose audio representations for both speech and non-speech-related tasks.
These studies sampled anchors, positives, and negatives from unlabeled data.
They used a metric learning method with triplet loss to minimize\,/\,maximize the distance between the anchor and positive\,/\,negative pairs.
More recently, COLA~\cite{cola} achieved higher performance, which is based on self-supervised learning with contrastive loss.
COLA samples positive pair as segments from the same audio clip and negative pair as segments from different clips, and maximize\,/\,minimize the similarity between positive\,/\,negative pairs based on multiclass cross-entropy.
COLA has been evaluated on multiple tasks, including speech, music, acoustic scenes, and animal sounds, and outperformed the performance of TRILL.
\begin{figure*}[tb]
  \centering
    \includegraphics[clip, width=17.8cm]{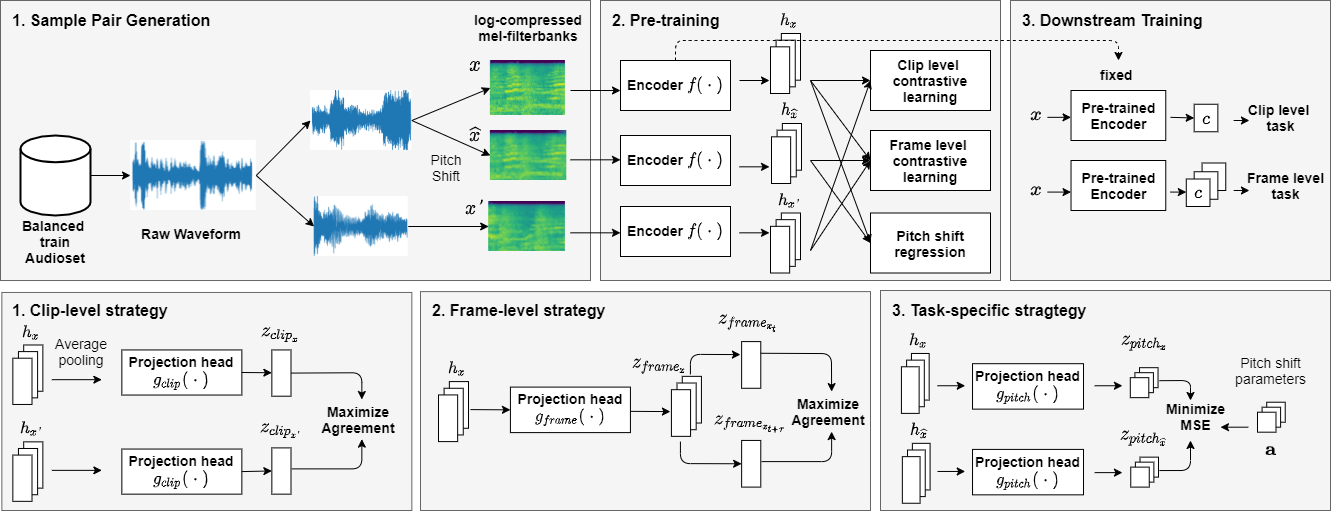}
  \caption{Overview of proposed method.}
  \label{fig:overview}
\end{figure*}

The key of training with contrastive loss is the sampling strategy, i.e., the definition of anchor, positive, and negative samples.
COLA assumes that data from the same audio clip is similar in time, so a pair of segments sampled from the same audio clip is a positive sample.
Different audio clips are assumed to have different characteristics from an anchor and are negative samples.
The sampling strategy is reasonable to the wide range of audio tasks like audio-tagging and speaker identification because they are clip-level classification.
The segments in clips should be the same classes in these tasks.
However, in some tasks, the samples do not necessarily belong to the same class even when they are highly similar in time. 
For example, SED requires frame-level classification of events; different segments in the same audio clip can not be the same class.
Also, it is necessary to identify the time changes from the same source in PD.
COLA is also unsuitable for the task because it maximizes the similarity between the pair of segments in the same clip, even when the pitch of the sound is different. 
So that we need to design the strategy carefully for each task, and it is difficult to learn general-purpose representations with a single sampling strategy.

We propose a self-supervised learning method using multiple sampling strategies for general-purpose audio representation.
The proposed method performs self-supervised learning with multiple contrastive losses based on multiple strategies.
By using multiple complementary strategies, a single model can be used for various tasks that the single conventional strategy cannot cover.
In this study, we introduce two new strategies, frame-level strategy and task-specific strategy, in addition to the clip-level sampling strategy of COLA, to improve the performance of frame-level classification and identify the time variation from the same source.

\section{Method}
\label{sec:method}
In this study, we perform self-supervised learning using three sampling strategies shown in Fig.~\ref{fig:overview}.
The proposed method is a multi-task learning method with three contrastive losses computed based on three different strategies.
The first strategy is a clip-level strategy that focuses on differences between audio clips. 
The second strategy is a frame-level strategy that focuses on time changes in audio segments.
The third strategy is a task-specific strategy that focuses on spectrum changes in a sound source. 
In the following, we discuss each strategy.

\subsection{Clip-level strategy}
\label{subsec:clip}
The first strategy is based on the clip-level sampling strategy used in COLA. 
The strategy defines a positive sample as audio segments from the same clip as an anchor segment, and negative segments are sampled from different audio clips.
COLA calculates the similarity between speech segments in two steps. 
First, encoder $f$ maps the acoustic feature data $x_{1:T}\in \mathbb{R}^{N\times T}$ into embedding vectors $h=f(x_{1:T})\in\mathbb{R}^{d\times T'}$, where $N$, $T$ and $T'$ are the number of frequency bins, time frames and the frame length of $h$, respectively.
Then, after global average pooling along with temporal frames, 
the shallow neural network $g_{clip}$ maps $h$ to the clip representations $z_{clip} = g_{clip}(h)\in\mathbb{R}^{d}$,
where bilinear comparisons are performed. 
The similarity of the representations of two segments $(x, x')$ can be expressed:
\begin{equation}
    \label{eq:s_clip}
    \begin{split}
        s_{clip}(x, x') = g_{clip}(f(x_{1:T}))^T W g_{clip}(f(x'_{1:T})),
    \end{split}
\end{equation}
where $W$ is the bilinear parameter, $x'$ is the feature value of the segment that is different from $x$. 
As an objective function, we rely on multiclass cross-entropy applied to the similarities:
\begin{equation}
    \label{eq:L_clip}
    \begin{split}
        \mathcal{L}_{clip} = -{\rm log}{\frac{{\rm exp}(s_{clip}(x, x^+))}{\sum{{\rm exp}(s_{clip}(x,x^-))}}},\\
        \scriptstyle
        x^-\in\mathcal{X}^-(x)\cup\{x^+\}
    \end{split}
\end{equation}
where $x^+$ is the positive associated to anchor $x$, $\mathcal{X}^-(x)$ refers to the set of negative distractors.
The embedding vectors $h$ obtained from the loss are effective for clip-level classification.
However, the embedding vectors are unsuitable for frame-level classification of events and identifying the time variation of the sound emitted from the same source.

\subsection{Frame-level strategy}
\label{subsec:frame}
The second strategy computes contrastive loss for frame-level embedding vectors based on the assumption that neighboring audio segments are similar and audio segments that are apart in time are dissimilar. 
The similarity for frame-level embedding vectors has two differences from the first clip-level strategy.
The first difference is that in \ref{subsec:clip}, cross-entropy is calculated as negative examples for all clips except for the same clip.
In \ref{subsec:frame}, cross-entropy is extended to calculate the embedding vectors of $T'' = \{\tau \,|\, 2m + 1 \cap 0 \leqq m \leqq \frac{T'}{2} \cap m \in \mathbb{Z}\}$ frames around the target frame as a positive example.
Second, while \ref{subsec:clip} calculates the cross-entropy for a mini-batch, \ref{subsec:frame} calculates the cross-entropy for the similarity of the embedding vectors of each sample and then calculates and averages the value for each sample of the mini-batch.

A shallow neural network $g_{frame}$ maps the embedding vectors $h$ to the frame representations $z_{frame} = g_{frame}(h)\in\mathbb{R}^{d\times T'}$.
The bilinear similarity of the frame representations of two segments $(x_t, x_{t+\tau})$ is calculated as:
\begin{equation}
    \label{eq:s_frame}
    \begin{split}
        s_{frame}(x_t, x_{t+\tau}) = g_{frame}(f(x_t))^T Wg_{frame}(f(x_{t+\tau})).
    \end{split}
\end{equation}
The objective function adapts the extended multiclass cross-entropy to the similarity of the frame representations:
\begin{equation}
    \label{eq:L_frame_x}
    \begin{split}
        \mathcal{L}_{frame}(x) = -\frac{1}{N'M}\sum_{n=1}^{N'}{\rm log}{\frac{\sum_{\tau}^{T''}{{\rm exp}(s_{frame}(x_t, x_{t+\tau}))}}{{\sum_{i=1}^{T'}{{\rm exp}(s_{frame}(x_t, x_i))}}}},
    \end{split}
\end{equation}
where $M$ is the number of elements in $T''$, and $N'$ is the number of elements in the mini-batch. 
In training, we use:
\begin{equation}
    \label{eq:L_frame}
    \begin{split}
        \mathcal{L}_{frame} = \frac{1}{3}\bigg\{\mathcal{L}_{frame}(x) + \mathcal{L}_{frame}(x') + \mathcal{L}_{frame}(\hat{x})\bigg\},
    \end{split}
\end{equation}
where $\hat{x}$ are the frame representations at the same time as $x$ transformed by pitch shift for $x$.
The embedding vectors $h$ are effective for frame-level classification.
\subsection{Task-specific strategy}
\label{subsec:pitch}
\ref{subsec:clip} and \ref{subsec:frame} assume that, despite differences in temporal resolution, temporal neighbors are sounds from the same source, i.e., similar embedding vectors.
However, for tasks such as PD, where changes in the sound source spectrum are important, it is not suitable to make different segments positive because even small changes in the spectrum in the temporal neighborhood can be an important discriminant for the task.
Our strategy is to artificially generate a positive sample for each segment by designing a new, unique data augmentation for each task.
In this study, we consider the problem of pitch-shift estimation.
We create a pitch-shifted anchor $\hat{x}$ by pitch-shifting the anchor $x$ with a shift width $a$ and detect how much the pitch-shifted anchor has been shifted for the anchor using the least-squares method.
We use a shallow neural network $g_{pitch}$ that maps the embedding vectors $h$ to a pitch shift detector, mapping it to the frame representations $z_{pitch} = g_{pitch}(h)\in\mathbb{R}^{T'}$. 
Since the pitch-shifted width is a relative measure, the loss function is calculated as follows:
\begin{equation}
    \label{eq:L_pitch}
    \begin{split}
        \mathcal{L}_{pitch} = \|g_{pitch}(f(\hat{x}_{1:T})) - g_{pitch}(f(x_{1:T})) - \mathbf{a}\|^2,
    \end{split}
\end{equation}
where $\mathbf{a}$ is a vector for length $T''$ that contains pitch shift parameter $a$. 
In this case, we considered data augmentation for pitch changes, but the data augmentation method and loss function should be changed for each task when applied to other problems.
The embedding vectors $h$ obtained from the loss are effective for PD.

\subsection{Final loss function}
\label{subsec:final_loss}
Finally, we combine the three loss functions for self-supervised learning:
\begin{equation}
    \label{eq:final_loss}
    \begin{split}
        \mathcal{L} = \mathcal{L}_{clip} + \alpha\mathcal{L}_{frame} + \beta\mathcal{L}_{pitch},
    \end{split}
\end{equation}
where $\alpha$ and $\beta$ are the hyperparameters.
\subsection{Transfer learning for downstream tasks}
The application to the downstream tasks is made in two steps: 1) the encoder $f$ of the pre-trained model is extracted and used as the feature extractor, 2) the feature extractor is frozen, and only the classifier is trained.

\section{Experiments}
\label{sec:exp}
The important point of the experiments is to determine if the pre-trained embedding vectors are adaptable across audio domains and recording conditions, not only for the clip classification task but also for SED and PD.
\subsection{Datasets and Tasks}
\label{subsec:datasets_and_tasks}
We pre-trained neural network models by the proposed method using a balanced subset of Audioset.
The dataset contains 18,939 samples of 10-second audio excerpts from YouTube videos.
Since our method is self-supervised, we do not use any labels. 
In this study, we used three types of tasks for evaluation. 
First, we used Google speech commands (SC) as a scene-based multi-class classification~\cite{speechcommandsv2}. 
The top one error was used for evaluation.
Second, we used DCASE 2016 for SED in synthesized sounds~\cite{Mesaros2018_TASLP,Lafay2017}. 
The evaluation followed the original DCASE and used the onset F-measure. 
Third, we used NSynth as a scene-based multi-class PD. 
The task classifies a standard MIDI piano sound into one of 88 pitches in a multi-class fashion~\cite{nsynth2017}.
As with CREPE, we evaluated it on pitch accuracy and chroma accuracy~\cite{crepe}.
\subsection{Experimental conditions}
\label{subsec:experimental_conditions}
Given an audio input sequence, a log-compressed mel spectrogram was extracted with a window size of 25\,ms,
a hop size of 10\,ms, and $N=64$\,mel spaced frequency bins in the range of 60\,-\,7800\,Hz corresponding to $T=96$\,frames, 960\,ms.
These features are passed to the encoder $f$ based on EfficientNet-B0~\cite{pmlr-v97-tan19a}, a lightweight and scalable convolutional neural network. 
Apply average pooling in the frequency direction of the last layer of the encoder to obtain embedding vectors $h$ of size $512\times3$.
This layer contains a 512-unit fully connected layer, followed by the Layer Normalization~\cite{ba2016layer} and tanh activation functions.
We trained the encoder by iterating 250 epochs for the subset of Audioset, using a batch size of 1024, a learning rate of $10^{-4}$, Adam~\cite{kingma2015}.
The hyperparameter $m$ of $T''$ was set to 0 due to $T''=3$.
$\alpha$ and $\beta$ in Eq.~\ref{eq:final_loss} was set to 1.0.
The shift width of the pitch shift was transformed every epoch to a random value in the range of 0.8 to 1.2.
In the experiment, we investigated the effect on each loss function and the effect of the similarity function.

For the downstream tasks, 
we trained the classifier by iterating 250 epochs for each dataset using a batch size of 1024, a learning rate of $10^{-4}$, Adam, and a dropout of 0.1, with a fully connected layer using the pre-computed embedding vectors as input.
We evaluated whether the embedding vectors trained by pre-training were universally adaptable regardless of the audio domain, recording conditions, or tasks.
\begin{table}[]
\centering
\caption{Test scores of a linear classifier trained on top of proposed embeddings or baseline pre-trained embeddings.}
\vspace{1mm}
\label{table:results1}
\scalebox{0.83}{\renewcommand\arraystretch{1.1}{$\displaystyle
\begin{tabular}{l|cccr}
\hline
 & \textbf{SC} & \textbf{DCASE} & \multicolumn{2}{c}{\textbf{NSynth}} \\
 & \textbf{acc} & \textbf{F1} & \textbf{pitch} & \multicolumn{1}{l}{\textbf{chroma}} \\ \hline\hline
\textbf{PANNs} & 0.083 & 0.754 & 0.012 & 0.096 \\\hline
\textbf{COLA} & 0.459 & 0.232 & 0.434 & 0.470 \\
$\mathcal{L}_{clip} + \mathcal{L}_{frame}$ & 0.535 & \textbf{0.344} & 0.424 & 0.462 \\
$\mathcal{L}_{clip} + \mathcal{L}_{pitch}$ & \textbf{0.585} & 0.244 & 0.428 & 0.463 \\
$\mathcal{L}_{clip} + \mathcal{L}_{frame} + \mathcal{L}_{pitch}$ & 0.572 & 0.278 & \textbf{0.452} & \textbf{0.487} \\ \hline
\end{tabular}
$}
}
\end{table}

\begin{table}[]
\centering
\caption{Test scores with different similarity functions.}
\vspace{1mm}
\label{table:sim_fc}
\scalebox{0.85}{\renewcommand\arraystretch{1.1}{$\displaystyle
\begin{tabular}{l|rrrr}
\hline
 & \multicolumn{1}{c}{\textbf{SC}} & \multicolumn{1}{c}{\textbf{DCASE}} & \multicolumn{2}{c}{\textbf{NSynth}} \\
 & \multicolumn{1}{c}{\textbf{acc}} & \multicolumn{1}{c}{\textbf{F1}} & \multicolumn{1}{c}{\textbf{pitch}} & \multicolumn{1}{c}{\textbf{chroma}} \\ \hline
\textbf{Cosine Similarity} & 0.401 & 0.094 & 0.212 & 0.238 \\
\textbf{Bilinear Similarity} & \textbf{0.572} & \textbf{0.278} & \textbf{0.452} & \textbf{0.487} \\ \hline
\end{tabular}
$}
}
\end{table}
\subsection{Results}
Table~\ref{table:results1} showed the scores in the three downstream tasks, using two baselines: the pre-trained model with Cnn14-DecisionLevelAtt in PANNs, and the model with COLA self-supervised trained on the subset of Audioset. 
We also compared the combination of loss functions.
First, we used the same procedure as in \cite{oord2019representation,He_2020_CVPR,tagliasacchi2019selfsupervised,pmlrv119chen20j} to evaluate the embedding vectors of the pre-trained proposed method with a linear classifier on the frozen embedding vectors. 
Compared to the other self-supervised learning methods, PANNs performed poorly on SC and NSynth, despite being trained on the full set of Audioset.
The result indicated that some tasks were not suitable for different learning methods. 
In addition, when comparing COLA with the proposed method, performance was improved in all tasks by adapting loss functions that combined three sampling strategies. 
The results showed that multiple perspectives allowed us to capture the same source's time variation and feature changes.
The ablation study of each loss function showed that combining $L_{frame}$ and $L_{pitch}$ improved the performance for NSynth.
In addition, the performance improvement is the largest when using $L_{frame}$ for SC and $L_{pitch}$ for DCASE, while the performance improvement is the smallest when using $L_{pitch}$ for SC and $L_{frame}$ for DCASE. 
There is a trade-off between $L_{frame}$ and $L_{pitch}$. 
Since $L_{pitch}$ was calculated at the frame-level, it can be considered a variant of frame-level loss.
There may be a trade-off between frame-level loss, which made the distance of the same source closer, and pitch loss, which results in different embedding when the pitch was different even for the same source.
These results showed that the combination of the three-loss functions was found to be important.

To investigate the role of the similarity function, we compared the pre-training of the model using cosine and bilinear similarity.
Cosine similarity was calculated as:
\begin{equation}
    \label{eq:cosine_sm}
    \begin{split}
        s(x, x') = \frac{g(f(x))^T\cdot g(f(x'))}{\|g(f(x))\|\|g(f(x'))\|},
    \end{split}
\end{equation}
where $g$ was each objective function's mapping function, $x$ and $x'$ were chosen appropriately for each objective function.
The calculated cosine similarity was normalized by dividing by 0.2.
Eq.~\ref{eq:cosine_sm} was used instead of Eq.~\ref{eq:s_clip} and Eq.~\ref{eq:s_frame}.
Table~\ref{table:sim_fc} showed the results.
The best results were obtained when bilinear similarity was used, indicating its effectiveness.

For future studies, we will conduct experiments with the full set of Audioset.
Experiments with the full set will provide a better comparison.
The extension of the proposed method to other modalities is also one of the important future tasks.
We believe that using multiple sampling strategies is also key to the tasks in the other modalities, such as audio-video.
A model trained without considering time information cannot be applied to tasks with frame-level classification, such as lip-reading or emotion recognition.
For example, in~\cite{chung2020}, they proposed two different loss functions for speaker verification and speech recognition. 
However, they trained two separate models for each task.
By designing multiple sampling strategies for other modalities, we can build more general pre-trained models.
\section{Conclusion}
\label{sec:conclusion}
This paper proposes a self-supervised learning method using multiple sampling strategies for general-purpose audio representation by designing loss functions using audio pairs obtained using multiple sampling strategies. 
Our method improves the performance of all tasks compared to existing methods in the experiment of the subset of Audioset.
The results show that it is effective to design and combine the loss functions according to the tasks.



\newpage
\section{References}
\printbibliography
\end{document}